\documentclass{article}[12pt]
\usepackage{amsfonts,inputenc,amssymb,hyperref,amsmath,graphicx,geometry,xcolor,cite}
\usepackage{authblk, comment}
\usepackage{mathrsfs}
\title{CFT reconstruction of local bulk operators in half-Minkowski space}

\author[$1$]{Arpan Bhattacharyya\footnote{\url{abhattacharyya@iitgn.ac.in}}}
\author[$2$]{Manas Dogra\footnote{\url{manasdogra2019@gmail.com}}}
\author[$2$]{Shubho R. Roy\footnote{\url{roy.shubho@gmail.com}}}

\affil[$1$]{\small{Indian Institute of Technology Gandhinagar, Gujarat-382355, India}}
\affil[$2$]{Department of Physics, Indian Institute of Technology, Hyderabad\\
Telengana 502285, India}

\date{}

\begin{document}

\maketitle

\begin{abstract}
We construct a holographic map that reconstructs massless fields (scalars, Maxwell field \& Fierz-Pauli field) in half-Minkowski spacetime in $d+1$ dimensions terms of smeared primary operators in a large $N$ factorizable CFT in $\mathbb{R}^{d-1,1}$ spacetime dimensions. This map is based on a Weyl (rescaling) transformation from the Poincar\'e wedge of AdS to the Minkowski half-space; and on the HKLL smearing function, which reconstructs local bulk operators in the Poincar\'e AdS in terms of smeared operators on the conformal boundary of the Poincar\'e wedge. The massless scalar field is reconstructed up to the level of two-point functions, while the Maxwell field and massless spin-2 fields are reconstructed at the level of the one-point function. We also discuss potential ways the map can be generalized to higher dimensions, and to the full Minkowski space.
\end{abstract}

\section{Introduction}

Maldacena's holographic duality or AdS/CFT correspondence \cite{Maldacena:1997re, Gubser:1998bc, Witten:1998qj, Aharony:1999ti}
has provided us with a nonperturbative formulation of quantum gravity in asymptotically Anti-de Sitter (AdS) spacetimes in terms of a nongravitational (large $N$ factorizable) conformal field theory (CFT). Equipped with this nonperturbative formulation, substantial progress has been made in understanding fundamental foundational aspects of quantum gravity, such as: how a quasilocal bulk spacetime emerges from the underlying CFT degrees of freedom, how the quantum entanglement dynamics of the CFT give rise to gravitational dynamics in bulk AdS, quantum properties of black holes including the black hole entropy puzzle, and even led to the prospects of a resolution of the black hole information paradox via unitary dynamics of the CFT. However, the situation is starkly different in backgrounds that are not asymptotically AdS. Although generic arguments from black hole physics suggest that, analogously to AdS/CFT, one must have holographic dual definitions or formulations of quantum gravity in generic spacetimes, e.g. asymptotically flat or de Sitter (dS) spacetimes, unlike AdS/CFT we do not have generic formulations of such holographic dualities apart from isolated examples \cite{Banks:1996vh, Anninos:2011ui}. Nevertheless, by examining the (asymptotic) symmetries of a given spacetime, it is possible to infer the global symmetries of the holographic dual (field) theory \cite{Sachs:1962zza, Brown:1986nw, Strominger:2001pn, Bondi:1962px, Sachs:1962wk, Barnich:2009se, Bagchi:2010zz,Barnich:2010eb,Compere:2013bya,Compere:2012jk,Afshar:2019axx}. Subsequently, from the information about the (global) symmetries of the dual field theory alone, a lot of the quantum gravitational dynamics in/of the bulk can be reconstructed or derived. As was done for AdS in the HKLL program \cite{Hamilton:2005ju, Hamilton:2006az, Hamilton:2006fh, Kabat:2011rz, Kabat:2012av, Kabat:2012hp, Kabat:2013wga}, purely using the generic constraints of conformal symmetries of the dual CFT, one can reconstruct dynamics quasilocal bulk fields in the large $N$ approximation. In this work our interest is in asymptotically flat spacetimes. There has been tremendous activity on holography in asymptotically flat spacetimes, especially in $3+1$-dimensions, over the past ten years, the so-called \emph{Celestial Holography} program (see \cite{Strominger:2013jfa,Strominger:2017zoo, Pasterski:2021raf, Raclariu:2021zjz,Prema:2021sjp,McLoughlin:2022ljp} for a review) whereby one attempt to reconstruct scattering matrices of interacting local bulk fields (including gravitons) in $\mathbb{R}^{3,1}$ out of correlators of conformal fields ($SL(2,\mathbb{C})$ primaries) supported on the so-called celestial 2-sphere $S^2$ obtained by suppressing the null direction of future null infinity. A second approach is the \emph{Carrollian Holography} program \cite{Bagchi:2010zz, Donnay:2022aba, Donnay:2022wvx, Bagchi:2022emh, Saha:2023hsl,Bagchi:2020fpr,Bagchi:2023fbj}\footnote{This list is by no means exhaustive. Interested readers are referred to the references and citations of these papers.} whereby the Poincar\'e group realised as conformal symmetries of (future) null infinity, $\mathscr{I}$ and the so-called Carrollian conformal fields supported on  $\mathscr{I}$ being dual to local fields in $\mathbb{R}^{3,1}$ (see \cite{Nguyen:2023vfz} for some details). For an alternative approach to flat space bulk reconstruction in terms of holography of information see \cite{Laddha:2020kvp}. However, in the case of the Celestial Holography or Carrollian Holography program, the boundary dual field theory is unknown. More generally, one must note that unlike in the case of the Maldacena duality, one does not know how to construct explicit gauge theory duals of asymptotically flat space gravity from top-down string compactifications \footnote{One can perhaps consider the $\mathcal{M}_3$/LST duality obtained by deforming a CFT$_2$ by irrelevant couplings \cite{Giveon:2017nie} as an exception, in the sense that the background metric is asymptotically flat. However, in this case, there is a non-trivial dilaton background that diverges at spatial infinity, and the dual theory (little string theory) is not a local field theory that does not admit UV completion.}. While the BFSS matrix model \cite{Banks:1996vh} formulation of M-theory in the lightcone frame does provide a dual quantum mechanics model of an asymptotically flat gravity background, the role of conformal infinity and asymptotic symmetries is not clear in this approach (see \cite{Miller:2022fvc, Tropper:2023fjr} for some recent attempts), so the conventional tools of holography such as the HKLL map etc cannot be utilized readily. Notwithstanding, one can try to gain insight into flat space gravity by constructing the smearing functions for flat space gravity either in the Celestial or Carrollian holography program by replicating the success of the HKLL program for asymptotically AdS gravity. In this work, we take a step towards this direction. We consider here half-Minkowski space, described by the metric in Cartesian coordinates $t,\boldsymbol{x},z$,
\begin{equation}
    ds^2 = -dt^2+d\boldsymbol{x}\cdot d\boldsymbol{x} + dz^2, 
\end{equation}
with a maximal range for $t,\boldsymbol{x}$, i.e. $t, \boldsymbol{x}\in (-\infty,\infty)$, while $z$ is allowed to take values on the half-line $z\in (0,\infty)$, instead of full Minkowski space. In some ways, the half-Minkowski space will be a counterpart of the Poincar\'e wedge of the global AdS space and are related to each other via a Weyl rescaling of the metrics. As such, the dynamics of local fields propagating in half-Minkowski space which transform nicely under Weyl rescaling transformations can be obtained from their dynamics in the Poincar\'e AdS wedge. But the dynamics of the local fields in the Poincar\'e wedge of AdS$_{d+1}$ can be obtained from their nonlocal holographic CFT dual operators living on the conformal boundary ($z=0$) of the Poincar\'e wedge namely Minkowski space $\mathbb{R}^{d-1,1}$, via the HKLL smearing function construction \cite{Hamilton:2006az, Hamilton:2006fh}. Putting these two facts together we obtain a holographic smearing function map for the local operators in the bulk of half-Minkowksi space, and nonlocal CFT operators in full Minkowski space with one less spatial dimension \footnote{In this regard, we should point out to the recent work \cite{Laddha:2022nmj} where the authors have reconstructed massive fields in full Minkowski space from observables at the spatial infinity.}. We note that this map, just like the HKLL map, is obtained in the planar limit $N\rightarrow \infty$, i.e. the leading semiclassical approximation in the quantum gravity in the bulk (vanishing Planck length). In this limit, the equations of motion of the local fields in the bulk are linear (free fields). One can, of course, perform $1/N$ corrections to this map and reconstruct the interacting physics of bulk fields in the half-Minkowski space as well, as was done for AdS \cite{Kabat:2011rz, Kabat:2012av, Kabat:2013wga}. However, we leave that exercise for follow-up works.
\\

The plan of the paper is as follows. In Sec. \ref{scalar}, we reconstruct the massless scalar propagating in half-Minkowski space by smeared primary operators in a CFT by first mapping the massless scalar to a conformally coupled scalar in the Poincar\'e wedge via a conformal transformation (Weyl rescaling) and then appropriating the HKLL map (smearing functions) \cite{Hamilton:2005ju,Hamilton:2006az,Hamilton:2006fh,Kabat:2012hp}. Our analysis is restricted to the case of CFT dimensions, $d=1,2$ for the sake of convenience: for these conformally coupled scalars, the HKLL smearing functions is not the correct one for $d\geq3$, and one should work with more general smearing constructions \cite{Aoki:2023lgr, Aoki:2021ekk, DelGrosso:2019gow}. After reconstructing the scalar field itself (one-point function) in Sec. \ref{CFT2flat smearing}, we proceed to reconstruct the Wightman functions in the half-Minkowski space in Sec. \ref{scalar 2pt reconstruction}. For $d=1$, we are able to reconstruct the bulk Wightman function (for both spacelike and timelike separations) analytically, while for $d=2$, the smearing integrals are performed numerically for the timelike separated case and shown to be equal to the bulk sugra Wightman function. In Sec. \ref{Maxwell4d}, we take up the reconstruction of the Maxwell field in $3+1$ dimensions for obvious phenomenological reasons. This is accomplished straightforwardly since the Maxwell equations are conformally invariant in $3+1$ dimensions. We end the section with comments on extending to higher spacetime dimensions. In Sec. \ref{Grav4d}, we look at linearized metric fluctuations in $3+1$ dimensional half-Minkowski spacetime. To this end, we begin with the equation for linearized metric fluctuation in the holographic gauge \cite{Kabat:2012hp} in the flat side and then perform a conformal transformation to the AdS Poincar\'e wedge. Conveniently this turns into the equation of motion for a free massless spin-2 field in AdS space with AdS radius scaled up by a factor of $\sqrt{3}$, again in the holographic gauge. This readily allows us to reconstruct the linearized metric fluctuation in the half-Minkowski space from the HKLL metric fluctuation reconstruction in holographic gauge \cite{Kabat:2012hp}. We end this section with comments on extending this construction to arbitrary spacetime dimensions. In Sec. \ref{Disc}, we discuss our results and provide an outlook for future work. Finally, in the appendices, we collect some useful mathematical details.

\section{CFT reconstruction of scalar fields in Half Minkowski space} \label{scalar}

The Poincar\'{e} wedge of AdS
\begin{equation} \label{met}
ds^2=\frac{l^2}{z^2} \left(dz^2+dx^\mu dx_\mu\right), z\geq 0
\end{equation}
 is conformal to half Minkowski space, with the conformal factor $\Omega^2(x)=l^2/z^2$. Consider a massless conformally coupled scalar in Poincar\'{e} AdS:
\begin{equation}
\left(\overline{\Box}-\frac{d-1}{4\,d}\overline{R} \right)\,\overline{\varphi}=0.\label{eq:KGAdS}
\end{equation}
where $\overline{\Box}$ is the d' Alembertian operator in  Poincar\'{e} AdS$_{d+1}$.
Since the Ricci scalar for $AdS_{d+1}$ is a constant, $R=-\frac{d(d+1)}{l^2}$ the equation of motion of the conformally coupled scalar turns into that of a minimally coupled massive scalar field, 
\begin{equation}
   \left(\overline{\Box}-\mu^2\right)\,\overline{\varphi}=0. 
\end{equation}
 of mass, 
 \begin{equation}
     \mu^2=-\frac{d^2-1}{4}. \label{mass}
 \end{equation}
 Note that this mass-squared is in the Breitenlohner-Freedman stability regime $\mu^2>-\frac{d^2}{4}$. This conformally coupled scalar is then dual to a quasi primary field in the CFT with conformal dimension \begin{equation}
    \Delta=\frac{d+1}{2}.\label{conformaldimension}
\end{equation}
Under a Weyl rescaling by the conformal factor, $\Omega^2(x)$, 
\begin{equation}
\overline{\varphi}=\Omega^{-\frac{d-1}{2}}\,\varphi \label{AdS2flat}
\end{equation}
the equation (\eqref{eq:KGAdS}) turns into a massless (conformally coupled) scalar field propagating in \emph{half}-Minkowski space,
\begin{equation}
\Box~\varphi=0.\label{eq:KGflat}
\end{equation}
Here $\Box$ is the d'Alembertian operator in $\mathbb{R}^{d,1}$. To see this we use the transformation properties of $\Box$ and the curvature scalar, $R$ under Weyl rescalings \cite{Wald:1984rg},

\begin{equation}
    \overline{\Box}~\overline{\varphi}=\Omega^{-\frac{d+3}{2}}\Box \varphi+\frac{1-d}{2} \Omega^{-\frac{d+5}{2}}(\varphi~\Box \Omega)-\frac{(d-1)(d-3)}{4}(\Omega^{-\frac{d+7}{2}}\varphi)|\nabla \Omega|^2\,,\end{equation}
\begin{equation}
    \frac{d-1}{4d}\overline{R}~\overline{{\varphi}}=\frac{d-1}{4d}\Omega^{-\frac{d+3}{2}} R \varphi+\frac{1-d}{2} \Omega^{-\frac{d+5}{2}}(\varphi~\Box \Omega)-\frac{(d-1)(d-3)}{4}(\Omega^{-\frac{d+7}{2}}\varphi)|\nabla \Omega|^2\,,
\end{equation}
where the unbarred quantities are in flat space, and the barred ones are in AdS space. Putting the above equations in \eqref{eq:KGAdS} along with the fact that $R=0$ in flat space we get \eqref{eq:KGflat}.\\

\subsection{The CFT to half-flat Smearing function} \label{CFT2flat smearing}

The conformally coupled scalar in Poincar\'e AdS can be reconstructed from the dual CFT defined over $\mathbb{R}^{1,d-1}$ using the HKLL smearing function \cite{Hamilton:2005ju, Hamilton:2006az,Hamilton:2006fh,Kabat:2012hp},
\begin{equation}
\overline{\varphi}(z,x) = \int d^d x'\, \overline{K}_\Delta (x,z|x')\, \mathcal{O}_\Delta(x')
\end{equation}
where 
\begin{equation}
\overline{K}_{\Delta} (x,z|x') = C(\Delta,d)\,\, \left(\sigma z'\right)^{\Delta-d}\:\theta(\sigma z') \label{HKLL1}
\end{equation}
where $C(\Delta, d)$ is a constant, which depends on the boundary spacetime dimensions and the conformal dimension of the primary\footnote{See App. \ref{HKLL redone} for a short derivation} after setting $\Delta = \frac{d+1}{2}$. . Then, upon performing the Weyl rescaling transformation \eqref{AdS2flat}, the massless scalar in the half-Minkowski space in $d,1$ dimensions can be reconstructed from the CFT in $\mathbb{R}^{1,d-1}$,
\begin{equation}
\varphi (z,x)= \int d^dx'\,K_\Delta (z,x|x')\, \mathcal{O}_\Delta (x')
\end{equation}
where,
\begin{equation}
K = \Omega^{\frac{d-1}{2}} \overline{K}_\Delta = \left(\frac{l}{z}\right)^{\frac{d-1}{2}} C(d)\,\left(\sigma z'\right)^{\Delta-d} \theta(\sigma z') \label{half flat smearing}
\end{equation}
where $C\left(\Delta=\frac{d+1}{2},d\right)\equiv C(d)$. This CFT to half-flat space smearing function for conformally coupled scalar fields, \eqref{half flat smearing},
can be expressed in Lorentz-invariant form as,
\begin{equation}
K\left(s\right)=2^{\frac{d-1}{2}}\,C(d)\, s^{-\frac{d-1}{2}}\:\theta(s)\label{eq: flat space HKLL for conformally coupled fields}
\end{equation}
where $s=z^{2}-\left(t-t'\right)^{2}-\left(\boldsymbol{y}-\boldsymbol{y}'\right)^{2}$
is the Lorentz invariant interval \emph{Wick-rotated} in the boundary spatial coordinates (i.e. $\boldsymbol{y}$
$\boldsymbol{y}\equiv \boldsymbol{x}$). One can directly
check (see App. \ref{flat space EOM check}) that this smearing function satisfies the equation of motion
for the conformally coupled field, namely
\[
\Box\:K(s)=0\,.
\]
However, since the HKLL smearing function CFT integral for a scalar is only well-defined for $\Delta>d-1$ \cite{Kabat:2012hp}, this flat-space smearing \eqref{half flat smearing} is well-defined only for $d<3$. For $d\geq3$, one has to work with a different smearing function (distribution), e.g. one that can be obtained using a mode sum approach.
\\
\subsection{CFT reconstruction of bulk Wightman functions} \label{scalar 2pt reconstruction}
 Using the universal form of the CFT two-point function, one has to show that \eqref{half flat smearing} reproduces the usual Wightman function for free massive scalar in flat space i.e. 
\begin{equation}
\langle {\varphi}(z_1,x_1)\, {\varphi}(z_2,x_2)\rangle  = \int d^dx'_1\,d^dx'_2\,\, {K}_\Delta(z_1,x_1|x'_1)\, K_\Delta(z_2,x_2|x'_2) \,\,\langle\mathcal{O}_\Delta (x'_1) \,\mathcal{O}_\Delta (x'_2) \rangle_{CFT{_d}} \label{2point}
\end{equation}
where 
\begin{equation}
    \langle\mathcal{O}_\Delta (x'_1) \,\mathcal{O}_\Delta (x'_2) \rangle_{CFT{_d}}=\frac{1}{(x_1'-x_2')^{2\Delta}}\,.
\end{equation}
Recall that we can have either $d=1$ or $d=2$; we take up these two cases separately.

\subsubsection{$d=1$ case}
For this case, the bulk SUGRA Wightman function is, 
\begin{align}
\langle \phi (z_1,t_1)\,\,\phi (z_2,t_2)\rangle &= \frac{1}{4\pi}\ln\left[\frac{(t_1-t_2)^2-(z_1+z_2)^2}{(t_1-t_2)^2-(z_1-z_2)^2}\right]\label{Wightman d=1} 
\end{align}
(see App. \ref{bulk Wightman}).\\\\
\textbf{CFT reconstruction:}\\
For $d=1$, $\Delta-d=0$ and the smearing function \eqref{half flat smearing} is simply the step function, restricting the support to the spacelike region. We proceed separately for two cases. First, we consider bulk operator insertions at spacelike separated points. For spacelike separated bulk points, we can choose $t_1=t_2=0$ and 
the right hand side of \eqref{2point} is
\begin{align}
\langle\phi (z_1,0)\,\phi (z_2,0)\rangle & = {C_1}^2 \int_{-{z_1}}^{z_1} dt_1' \int_{-{z_2}}^{z_2} dt_2' \frac{1}{(t_1'-t_2' - i \varepsilon )^{2}}\,, \nonumber \\&= {C_1}^2 \int_{-{z_1}}^{z_1} dt_1' \left(\frac{1}{t_1'-z_2}-\frac{1}{t_1'+z_2'}\right)\,, \nonumber %\\ &=\ln(z_1-z_2)-\ln(-z_1-z_2)-\ln(z_1+z_2)+\ln(-z_1+z_2) \nonumber
\\&= {C_1}^2 \ln\frac{(z_1-z_2)^2}{(z_1+z_2)^2}.
\end{align}
Next we consider the case when the bulk operators locations are timelike separated. To further simplify the situation, we choose $t_2=0$ and $z_1=z_2=z$ (note that the condition for no overlap over the boundary smearing supports is $t_1~\ge~2z$).
\begin{align}\label{2pttimelike}
    \langle \phi(t_1,z)\phi(0,z) \rangle&= {C_1}^2 \int_{t_1-z}^{t_1+z} dt_1' \int_{-z}^z dt_2'~\frac{1}{(t_1'-t_2'-i\varepsilon)^2}\,, \nonumber\\
    &= {C_1}^2 \left[\int_{t_1-z}^{t_1+z}\frac{dt_1'}{t_1'-z+i\varepsilon}-\int_{t_1-z}^{t_1+z}\frac{dt_1'}{t_1'+z+i\varepsilon} \right] \,,\nonumber\\
    &= {C_1}^2 \left[\ln(t_1-i\varepsilon)-\ln(t_1-2z-i\varepsilon)-\ln(t_1+2z-i\varepsilon)+\ln(t_1-i\varepsilon)\right]\,, \nonumber \\
    &= {C_1}^2 \ln \frac{(t_1-i\varepsilon)^2}{(t_1-i\varepsilon)^2-4z^2}\,.
\end{align}
Thus in both cases, we recover the SUGRA result \eqref{Wightman} if we identify the factor $ {C_1}^2=-\frac{1}{4\pi}$. The sign discrepancy arises from the fact that we are using the ``mostly minus" metric convention for field theory (CFT) while on the SUGRA side, we are using the ``mostly plus" convention appropriate to GR. The factor of $4\pi$ is due to the difference in normalizing the coefficient of the two-point function in CFT (where it is $1$) versus the usual SUGRA normalization of the two-point function. This $4\pi$ can be easily fixed by inserting a factor $1/\sqrt{4\pi}$ in the smearing function \eqref{half flat smearing}.\\
\subsubsection{$d=2$ case}
\begin{figure}[htb]
    \centering
    \includegraphics[width=\textwidth]{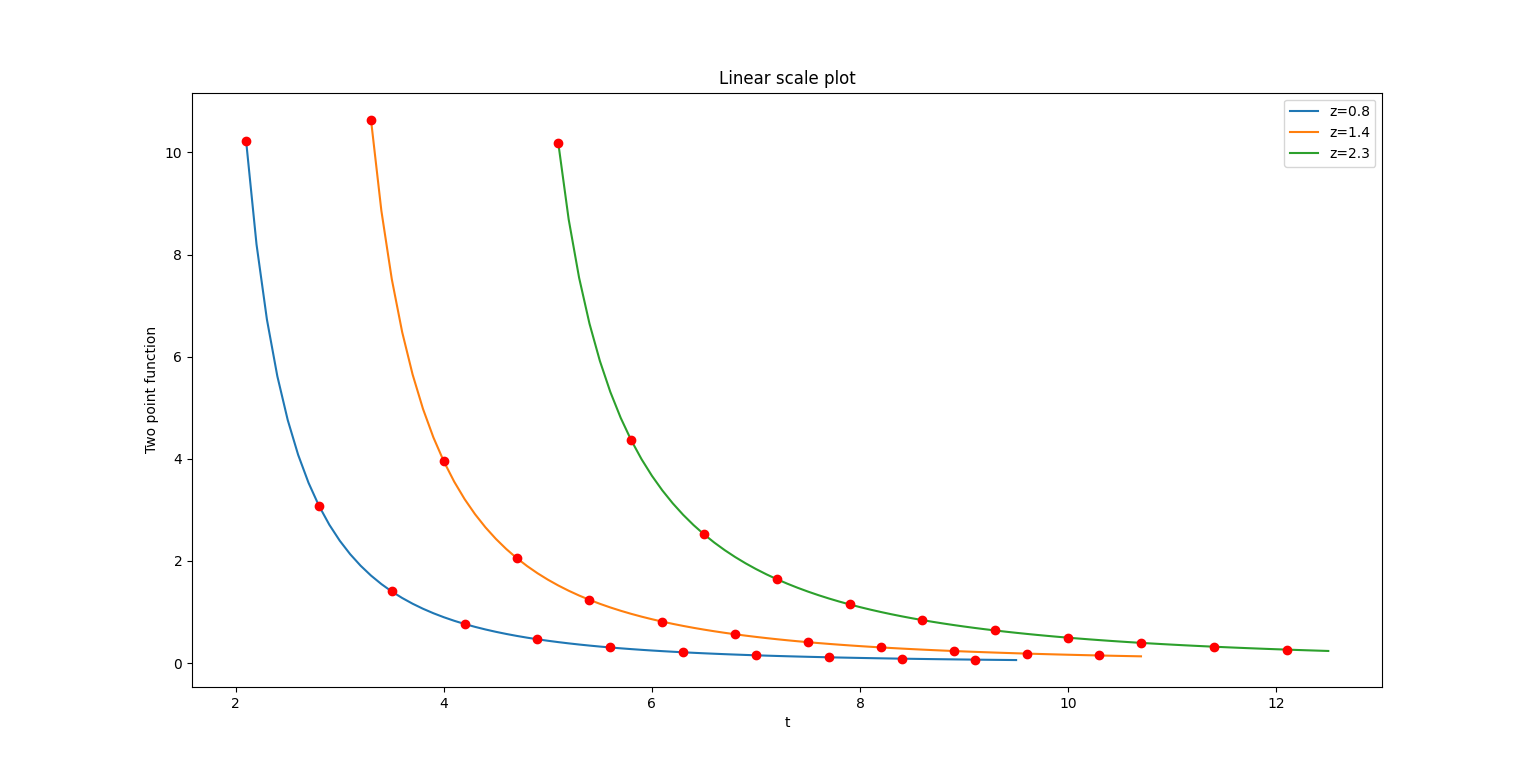}
     \caption{{\small\ 2 point function as a function of $t$ for different values of z }}\label{Figure_1}
\end{figure}

The half-Minkowski space (bulk) Wightman function when $d=2$ is 
\begin{equation}
   \langle \varphi(X)\,\varphi(Y)\rangle=\frac{1}{4\pi} \left(\frac{1}{|X-Y|}+\frac{1}{|X-Y_{*}|}\right) 
\end{equation}
where $X=(t_1,x_1,z_1)$, $Y=(t_2,x_2,z_2)$ are the location of two local operators and $Y_*=(t_2,-x_1,-z_2)$ is the location of the image operator. As is usual for the Wightman function, one needs to introduce an $i \varepsilon$ prescription by replacing $t_1 \rightarrow t_1+i\varepsilon$. Refer to App. \ref{bulk Wightman} for the details. We will be content with reconstructing the case for which $X,Y$ are timelike separated, $X=(t_1,0,z)$ and $Y=(t_2,0,z)$ with $t_1-t_2>2z$ so that there is no need for the $i \varepsilon$ prescription.\\\\
\textbf{CFT reconstruction:} Noting that $\Delta=\frac{3}{2}$ for $d=2$ from \eqref{conformaldimension}, and using \eqref{half flat smearing} and \eqref{2point}, the CFT integral representation of the bulk Wightman function for $d=2$ is,
\begin{equation}
   \langle \varphi(t_1,0,z_1) \varphi(t_2,0,z_2))\rangle = 2\,l\,{C_2}^2 \int dx'_1 \,dx'_2\,dt'_1\,dt'_2\, \frac{\left[(-(t_1'-t_1)^2-x_1'^2+z_1^2)(-(t_2'-t_2)^2-x_2'^2+z_2^2)\right]^{-\frac{1}{2}}}{[-(t_1'-t_2')^2-(x_1'-x_2')^2]^{\frac{3}{2}}}.
\end{equation}
The range of the integrals is determined by,
\[
(t'_1-t_1)^2 +{{x'}_1}^2 \leq {z_1}^2,\,\,\,(t'_2-t_2)^2 +{{x'}_2}^2 \leq {z_2}^2.
\]
Here we have set $x_1=x_2=0$. For timelike separated points, we can conveniently choose $z_1=z_2=z$ and $t_2=0$, and additionally, we can set $t_1>2z$ so that the two smeared operators in the CFT do not overlap. In such case, the above expression becomes,
\begin{align*}
   \langle \varphi(t_1,0,z)\,\varphi(0,0,z)\rangle & =2\,l\,C_2^2 \int_{(t'_1-t_1)^2 +{x'}^2_1\leq {z}^2}  dx'_1  dt_1' 
    \int_{t'^2_2 +x'^2_2 \leq z^2} dx'_2 dt'_2 \\
    & \qquad\qquad\qquad\qquad\qquad\qquad\qquad\qquad\qquad\qquad \frac{\left[(-(t_1'-t_1)^2-x_1'^2+z^2)(-t_2'^2-x_2'^2+z^2)\right]^{-\frac{1}{2}}}{[-(t_1'-t_2')^2-(x_1'-x_2')^2]^{\frac{3}{2}}}.
\end{align*}
To determine the constant $\alpha$, we perform the integrals for small $z$ and compare with the small $z$ expansion of the bulk SUGRA Wightman function \eqref{eq: Wightman function for t-separated points in 1/2 Minkowski} (see App. \ref{fix C}). We obtain \begin{equation}
   C_2=\frac{1}{4\pi^{3/2} l^{1/2}}.
\end{equation}
Thus the appropriately normalized (SUGRA norm) smeared CFT two-point function becomes
\begin{align}
    \langle \varphi(t_1,0,z)\,\varphi(0,0,z)\rangle & = \frac{1}{8\pi^3} \int_{-z}^z dx'_1  
    \int_{t_1-\sqrt{z^2-x_1'^2}}^{t_1+\sqrt{z^2-x_1'^2}} dt_1' 
    \int_{-z}^z dx'_2 
    \int_{-\sqrt{z^2-x_2'^2}}^{\sqrt{z^2-x_2'^2}} dt'_2 \nonumber \\
    & \qquad\qquad\qquad\qquad\qquad\qquad\frac{\left[(-(t_1'-t_1)^2-x_1'^2+z^2)(-t_2'^2-x_2'^2+z^2)\right]^{-\frac{1}{2}}}{[-(t_1'-t_2')^2-(x_1'-x_2')^2]^{\frac{3}{2}}}. \label{smeared2pt}
\end{align}
This integral is not tractable analytically and we proceed further numerically. The integral is computed numerically for three distinct values of $z$, and for many different values $t_1$ for each $z$. The results are shown graphically in Fig. \ref{Figure_1}, where the CFT integral representation of the bulk two-point function \eqref{smeared2pt} is plotted as a function of $t_1$.
The solid curves in the correspond to the bulk result \eqref{eq: Wightman function for t-separated points in 1/2 Minkowski} while the points are the numerically computed values of the CFT smearing integral \eqref{smeared2pt}. \begin{figure}[htb]
    \centering
    \includegraphics[width=\textwidth]{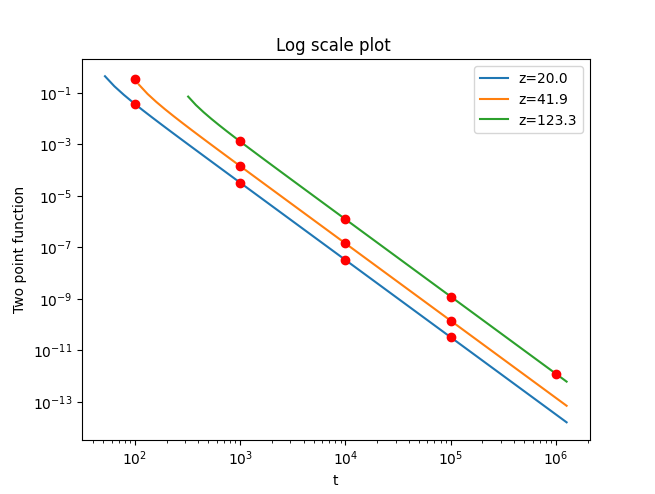}
     \caption{{\small\ Log-Log plot of 2 point function as a function of $t$ for high values of $z$ }}\label{Figure_2}
\end{figure} Fig. \ref{Figure_2}) shows the log-log plot version of the same data (solid curves represent the bulk Wightman function, while the points represent numerically computed values of the \eqref{smeared2pt}). From the numerical estimates, it is obvious the smearing of the boundary (CFT$_2$) Wightman function \eqref{smeared2pt} reconstructs that bulk Wightman function \eqref{eq: Wightman function for t-separated points in 1/2 Minkowski} for the timelike separated case (with non-overlapping smearing regions, $t_1>2z$). The spacelike separated case is a bit more subtle as one has to keep an $i \varepsilon$ prescription in the CFT two-point function due to the overlapping of the smeared regions along the time direction. We leave this case for future work.\\

\section{Maxwell field reconstruction}\label{Maxwell4d}

Now we discuss the CFT reconstruction of a Maxwell field in half-Minkowski spacetime, specifically in $3+1$ dimensions. We start with the AdS$_4$ metric in Poincar\'e  patch. We remind ourselves that 
%\begin{equation}
 %   ds^2=\frac{dz^2-dt^2+dx^2+dy^2}{z^2}, \quad z \geq 0.
%\end{equation}
that this metric is conformal to Minkowski half-space with the conformal factor is defined in the Sec.~(\ref{scalar}). 
%We can write the following,
%\begin{equation}
 %   g_{\textrm{h-mink}}= \Omega^2 g_{AdS}, \quad \Omega= z. \label{conftrans}
%\end{equation}
%$\textrm{h-mink}$ denote half of the Minkowski spacetime as $z\geq 0.$ 
Now we know that Maxwell theory, defined by the action, 
\begin{equation}
    S=-\frac{1}{4}\int d^4 x\, F^{\mu\nu}F_{\mu\nu},
\end{equation}
remains invariant under the conformal transformation of the form mentioned below (\ref{met}) in $3+1$ spacetime dimensions. It can be also shown that, 
$F_{\mu\nu}$ hence $A_{\mu}$ remains invariant under this scale transformation in $3+1$ dimensions. In \cite{Kabat:2012hp}, the reconstruction of the Maxwell gauge field have been done for Poincar\'e AdS$_d$. Given the fact that in $3+1$ dimensions gauge field remains invariant under conformal transformation, it straightforwardly follows that  we can  reconstruct the bulk Maxwell gauge field for the Minkowski half-spacetime and the kernel will be \emph{identical} to that given in \cite{Kabat:2012hp} provided one use the same gauge as \cite{Kabat:2012hp}, namely the \emph{holographic gauge} where we choose
\begin{equation}
A_z(z,x)=0, \partial^\mu A_\mu = 0.
\end{equation} 
We should note that, here, the CFT data is given at $z=0$ hyperplane, which is not the null infinity of Minkowski space. It is also obvious that the reconstruction of the Maxwell two-point function using CFT smeared operators will proceed identically to \cite{Kabat:2012hp} in a holographic gauge. As observed for the scalar case, if one projects out the pieces of the 2-point function (in general $n$-point function) which are not invariant under translation symmetry, then one obtains a reconstruct the Maxwell field for the $entire$ Minkowski spacetime from a CFT! \par
Note that the Maxwell theory is invariant under scale transformation other than $3+1$ dimensions. So the argument presented here will not hold other than $3+1$ dimensions. However, as will become evident from the CFT reconstruction of graviton fields in the next section, there is a possible way to deal with the CFT reconstruction of Maxwell fields in general spacetime dimensions even without the conformal invariance of Maxwell action.

\section{Reconstruction of bulk metric fluctuations}\label{Grav4d}

Finally, we will discuss the possibility of reconstructing the graviton field for half-Minkowski spacetime from CFT. Note that we want to reconstruct the graviton field, which satisfies the $3+1$ dimensional Einstein's equations. First, we note the following property of the linearized massless spin-2 field under Weyl rescalings, 
\begin{equation}
    h_{\mu\nu}=\Omega^2\, \tilde h_{\mu\nu}, \quad \Omega=z.\label{conftrans1}
\end{equation} 
We have set the AdS radius to unity at this point for convenience. We will restore it in the final step. 
Here $h_{\mu\nu}$ satisfies the linearized Einstein equation in $3+1$ dimensional half-Minkowski spacetime while $\overline{h}$ is the linearized metric perturbation in Poincar\'e AdS (but with a different cosmological constant, refer to \eqref{eq1}). To be precise, $\tilde h_{\mu\nu}$ satisfies the corresponding linearized equation obtained by Weyl transforming the linearized Einstein equation in flat space to the AdS$_4$ spacetime in the Poincar\'e patch. In $3+1$ dimensional (half)-Minkowski spacetime, the linearized Einstein equation can be found from the linearized Ricci flat equation, i.e. 
\begin{equation}
    R_{\mu\nu}=0. 
\end{equation}
Under the conformal transformation (\ref{conftrans1}) for generic metric $g_{\mu\nu}=\Omega^2 \tilde g_{\mu\nu}$ we get, 
\begin{equation}
    R_{\mu\nu}= \tilde{R}_{\mu\nu}- \tilde g_{\mu\nu} \tilde \nabla^2\ln (\Omega)+(d-2)\Big(-\tilde \nabla_\mu\partial_{\nu}\ln (\Omega)+\partial_{\mu}\ln (\Omega)\partial_{\nu}\ln (\Omega)-\tilde g_{\mu\nu}\partial_{\alpha}\ln (\Omega)\partial^{\alpha}\ln (\Omega)\Big), \label{ricci}
\end{equation}
where $d$ is the spacetime dimensions. Now we need to linearize it, i.e.
\begin{equation}
    g_{\mu\nu}^{(0)}+h_{\mu\nu}=\Omega^2\Big(\tilde g_{\mu\nu}^{(0)}+ \tilde h_{\mu\nu}\Big).
\end{equation}
We will eventually consider the background metrics $g_{\mu\nu}^{(0)}$ and $\tilde g_{\mu\nu}^{(0)}$ are half-Minkowski and AdS metric respectively. First of all, we note that by linearizing the left-hand side of (\ref{ricci}) and setting to zero we get the usual linearized Einstein equation for $h_{\mu\nu}$ i.e 
\begin{equation}
   \frac{1}{2}\nabla^2 h_{\mu\nu}+ \nabla_{(\mu}v_{\nu)}=0, \quad v_{\nu}=\frac{1}{2}\partial_{\nu}h-\nabla_{\alpha}h^{\alpha}{}_{\nu}.
\end{equation}
$h$ is the trace of $h_{\mu\nu}.$ Also, we have set $d=4.$
We can further simplify this to get,
\begin{equation}
    \partial^2 h_{\mu\nu}-\partial_{\alpha}(\partial_{\mu}h^{\alpha}{}_{\nu}+\partial_{\nu}h^{\alpha}{}_{\mu})+2 \partial_{\mu}\partial_{\nu}h=0.
\end{equation}
This is the usual wave equation for graviton on the (half)-Minkowski spacetime.\par

We also linearize the right-hand side of (\ref{ricci}). From the zeroeth order term we get,
\begin{equation}
\tilde R^{(0)}_{\mu\nu}-\tilde g^{(0)}_{\mu\nu}\tilde (\nabla^{(0)})^2 \ln(\Omega)+2 \Big(-\tilde \nabla_\mu^{(0)}\partial_\nu \ln(\Omega)+(\partial_\mu \ln(\Omega))(\partial_\nu \ln (\Omega))-\tilde g^{(0)}_{\mu\nu} (\partial\ln(\Omega))^2\Big)=0.
    \end{equation}
    This indeed gives the following equation in $3+1$ dimensions for $\Omega=z$, 
    \begin{equation}
        \tilde R^{(0)}_{\mu\nu}=\Lambda
        \tilde g^{(0)}_{\mu\nu},
    \end{equation}
    $\Lambda$ is the 4 dimensional cosmological constant. $$\Lambda=-3.$$
    This is exactly what the AdS$_4$ metric should satisfy. Next we look at the linearized equation for the perturbation $\tilde h_{\mu\nu}.$ We get the following,
    \begin{align}
    \begin{split}
 &  \frac{1}{2}\tilde\nabla_{\alpha}^{(0)}\Big(\tilde \nabla^{(0)}_\mu h^{\alpha}{}_{\nu}+\tilde \nabla^{(0)}_\nu h^{\alpha}{}_{\mu}\Big)-\frac{1}{2}\tilde \nabla_{\mu}\partial_{\nu}h-\frac{1}{2}(\tilde\nabla^{(0)})^2 h_{\mu\nu}  -h_{\mu\nu}(\tilde \nabla^{(0)})^2\ln(\Omega)\\& +\Big(\tilde g_{\mu\nu}^{(0)}\tilde \nabla^{(0)\beta}h^{\alpha}{}_\beta-\frac{1}{2}\tilde g_{\mu\nu}^{(0)}\tilde \nabla^{(0)\alpha}h+\tilde \nabla^{(0)}_{\mu}h^{\alpha}{}_{\nu}+\tilde \nabla^{(0)}_{\nu}h^{\alpha}{}_{\mu}\Big)\partial_{\alpha}\ln(\Omega)\\&
  +\tilde g_{\mu\nu}^{(0)}h^{\alpha\beta}\tilde \nabla_{\alpha}^{(0)}\tilde \nabla_{\beta}^{(0)}\ln(\Omega)-2\Big(h_{\mu\nu}(\partial\ln(\Omega))^2-\tilde g_{\mu\nu}^{(0)}h^{\alpha\beta}\partial_{\alpha}\ln(\Omega)\partial_{\beta}\ln(\Omega)\Big).
       \end{split} \label{linequa}
       \end{align}
     Noting that $$  (\tilde \nabla^{(0)})^2\ln(\Omega)=-3,$$ we can see that first line of (\ref{linequa}) is coming from linearizing the Ricci equation on AdS background i.e linearizing the left hand side of the equation below, 
     \begin{equation}
     R_{\mu\nu}+3 g_{\mu\nu}=0.
     \end{equation}
     Next following \cite{Kabat:2012hp}, we choose the holographic gauge i.e $$h_{zz}=0=h^z{}_{z}, h_{z\mu}=0=h^z{}_{\mu}, h^{a}{}_{a}=0$$ where, $a$ takes three values $t,x,y.$ 
     %Then we note several points, 
   %  \begin{itemize}
      %   \item $g_{\mu\nu}^{(0)}h^{\alpha\beta}\partial_{\alpha}\ln(\Omega)\partial_{\beta}\ln(\Omega)=g_{\mu\nu}^{(0)}h^{zz}\partial_{z}\ln(\Omega)\partial_{z}\ln(\Omega)=0$ as $h_zz$ and hence $h^{zz}$is zero. The indices are raised and lowered by the background metric $\tilde g^{(0)}$ and $\Omega=z.$
      %   \item \begin{align}\begin{split} &\Big(\tilde g_{\mu\nu}^{(0)}\tilde \nabla^{(0)\beta}h^{\alpha}{}_\beta-\frac{1}{2}\tilde g_{\mu\nu}^{(0)}\tilde \nabla^{(0)\alpha}h+\tilde \nabla^{(0)}_{\mu}h^{\alpha}{}_{\nu}+\tilde \nabla^{(0)}_{\nu}h^{\alpha}{}_{\mu}\Big)\partial_{\alpha}\ln(\Omega)\\&=
       %  \Big(\tilde g_{\mu\nu}^{(0)}\tilde \nabla^{(0)\beta}h^{z}{}_\beta-\frac{1}{2}\tilde g_{\mu\nu}^{(0)}\tilde \nabla^{(0)z}h+\tilde \nabla^{(0)}_{\mu}h^{z}{}_{\nu}+\tilde \nabla^{(0)}_{\nu}h^{z}{}_{\mu}\Big)\partial_{z}\ln(\Omega)
        % \end{split}\end{align}
         Note that, $h=h^{z}{}_{z}+h^{a}{}_{a}.$ By the choice of the gauge $h^{z}{}_{z}=0$ and $h^{a}{}_{a}=0.$ Hence $h=0.$ Also, the derivatives are w.r.t background metric hence i.e $\tilde \nabla^{(0) \alpha} \tilde g_{\gamma\beta}^{(0)}=0.$ 
         %Hence $\Big(\tilde g_{\mu\nu}^{(0)}\tilde \nabla^{(0)\beta}h^{\alpha}{}_\beta-\frac{1}{2}\tilde g_{\mu\nu}^{(0)}\tilde \nabla^{(0)\alpha}h+\tilde \nabla^{(0)}_{\mu}h^{\alpha}{}_{\nu}+\tilde \nabla^{(0)}_{\nu}h^{\alpha}{}_{\mu}\Big)\partial_{\alpha}\ln(\Omega)=0.$
       %  \item Last but not the least, 
       %  \begin{align}
      %       \begin{split}
     %            &h^{\alpha\beta}\tilde \nabla_{\alpha}^{(0)}\tilde \nabla_{\beta}^{(0)}\ln(\Omega)\\&=h^{zz}\tilde \nabla_{z}^{(0)}\tilde \nabla_{z}^{(0)}\ln(\Omega)+h^{ab}\tilde \nabla_{a}^{(0)}\tilde \nabla_{b}^{(0)}\ln(\Omega)+h^{za}\tilde \nabla_{z}^{(0)}\tilde \nabla_{a}^{(0)}\ln(\Omega).
    %         \end{split}
   %        \end{align}
    %        We can easily check that, $\tilde \nabla_{z}^{(0)}\tilde \nabla_{a}^{(0)}\ln(\Omega)=\tilde \nabla_{z}^{(0)}\tilde \nabla_{z}^{(0)}\ln(\Omega)=0$ and $\tilde \nabla_{a}^{(0)}\tilde \nabla_{b}^{(0)}\ln(\Omega)=-\frac{\eta_{ab}}{z^2}.$
     %        Then we get, 
 %\begin{align}
   %          \begin{split}
  %             &h^{zz}\tilde \nabla_{z}^{(0)}\tilde \nabla_{z}^{(0)}\ln(\Omega)+h^{ab}\tilde \nabla_{a}^{(0)}\tilde \nabla_{b}^{(0)}\ln(\Omega)+h^{za}\tilde \nabla_{z}^{(0)}\tilde \nabla_{a}^{(0)}\ln(\Omega)\\&=-\frac{1}{z^2}h^{ab}\eta_{ab}=-\frac{1}{z^2}h^{a}{}_{a}=0.
  %           \end{split}
 %           \end{align}
    % \end{itemize}
     So in this gauge, only non-vanishing contribution to (\ref{linequa}) comes from the following terms,
     \begin{align}
         \begin{split}
           \frac{1}{2}\tilde\nabla_{\alpha}^{(0)}\Big(\tilde \nabla^{(0)}_\mu h^{\alpha}{}_{\nu}+\tilde \nabla^{(0)}_\nu h^{\alpha}{}_{\mu}\Big)-\frac{1}{2}\tilde \nabla_{\mu}\partial_{\nu}h-\frac{1}{2}(\tilde\nabla^{(0)})^2 h_{\mu\nu}  -h_{\mu\nu}(\tilde \nabla^{(0)})^2\ln(\Omega)-2 h_{\mu\nu}(\partial\ln(\Omega))^2.  
         \end{split}
     \end{align}
     Now $(\nabla^{(0)})^2\ln(\Omega)=-3$ and $(\partial\ln(\Omega))^2=1.$ So we get 
     \begin{align}
         \begin{split}
           \frac{1}{2}\tilde\nabla_{\alpha}^{(0)}\Big(\tilde \nabla^{(0)}_\mu h^{\alpha}{}_{\nu}+\tilde \nabla^{(0)}_\nu h^{\alpha}{}_{\mu}\Big)-\frac{1}{2}\tilde \nabla_{\mu}\partial_{\nu}h-\frac{1}{2}\tilde\nabla^{(0)} h_{\mu\nu} + h_{\mu\nu}.  
         \end{split}
     \end{align}
     The first four terms come from linearizing the Ricci tensor. So this entire expression comes from varying left hand-side of the following Ricci equation, 
     \begin{align}
         R_{\mu\nu}+g_{\mu\nu}=0\,.
     \end{align}
     Restoring the AdS length scale ($l$) we get,
     \begin{align} \label{eq1}
         R_{\mu\nu}+\frac{1}{l^2}g_{\mu\nu}=0\,.
     \end{align}
     Note that the AdS in $3+1$ dimensions satisfies the following equation:
     \begin{align} \label{eq2}
         R_{\mu\nu}+\frac{3}{l^2}g_{\mu\nu}=0\,.
     \end{align}
     The perturbation in $3+1$ dimensional spacetime satisfies a similar equation to (\ref{eq2}) after a conformal transformation but with length scale scaled by a factor $\sqrt{3}$, i.e. $ \rightarrow \sqrt{3} l$ as evident from (\ref{eq1}).

The linearized metric perturbations of \eqref{eq1} can be reconstructed using the holographic gauge HKLL smearing functions \cite{Kabat:2012hp} from the CFT$_3$ stress tensor. Schematically,
\begin{equation}
    h_{ab}  \left(z,\boldsymbol{x}\right)= \int d^3 x\ K(z,\boldsymbol{x}| \boldsymbol{y})\, T_{ab} \left(\boldsymbol{y}\right),
\end{equation}
where $\boldsymbol{y}$ is the Wick-rotated boundary coordinate. In this section, the Greek indices $\mu, \nu$ denote the bulk spacetime components and the Latin indices $a,b$ denote the CFT indices. Finally, one can perform a reverse Weyl rescaling to go back to the original half-flat space, thereby completing the bulk reconstruction of the half-flat space metric perturbation in terms of the smeared stress-tensor of the CFT$_3$.

\begin{comment}
\textcolor{red}{Add concluding comments regarding reconstruction of one-point function and two-point function and Maxwell case other than 3+1 dimensions.}
\end{comment}

\section{Discussions and Outlook}\label{Disc}

In this work, we constructed smearing functions (extrapolate dictionary) representing local operators in half-Minkowksi space as nonlocal operators in a CFT with one less spatial dimension. The bulk fields reconstructed here, namely massless (conformally coupled) scalar, Maxwell field and Fierz-Pauli field (linearized metric perturbations) all have nice transformation properties under bulk conformal transformations, and we exploited this property to first map them to corresponding fields in the Poincar\'e wedge of AdS and then use the bulk reconstruction program for Poincar\'e AdS \cite{Hamilton:2006az, Kabat:2012hp}. This is an alternative to traditional holographic bulk reconstruction whereby one reconstructs bulk fields in terms of nonlocal operators on the conformal boundary of the space. Instead, here the bulk (half-Minkowski) spacetime possesses a physical boundary at $z=0$ (a ``Dirichlet" screen, since the bulk fields vanish here) and the holographic dual CFT degrees of freedom are localized on this screen. In this regard, our half-flat space holographic construction can be thought of as a alternative to the celestial or Carrollian holographic constructions. Our set up can be contrasted with the type of flat space bulk holography one can attempt to obtain by taking a large AdS radius (flat limit) of AdS/CFT \cite{Polchinski:1999ry}. We do not directly send the AdS radius to infinity here but instead use a Weyl rescaling transformation to get rid of the AdS-metric conformal factor in Poincar\'e coordinates. Our construction was performed entirely in the leading supergravity approximation (large $N$, large $\lambda$) where the bulk fields are free. We verified that the smearing function can be used to reconstruct massless conformally coupled scalar fields both at the one-point function and the two-point function (Wightman) level (for timelike separated local operator insertions). For the case of the Maxwell field, we restricted ourselves to $3+1$-dimensions where the Maxwell equations remains invariant under conformal transformations and we operated entirely in the holographic gauge \cite{Kabat:2012hp}. We verified that the Maxwell equations in holographic gauge in flat space in holographic are indeed obeyed by the smeared CFT$_3$ operators proposed. Finally we considered the linearized metric perturbations (Fierz-Pauli field) again in $3+1$-dimensions for phenomenological reasons. Although the (fully nonlinear) Einstein field equations in $3+1$-dimensions are not Weyl invariant, we demonstrated that the under the Weyl rescaling transformation (which takes half-flat space to Poincar\'e AdS), the flat space Einstein field equations transform to the field equations in AdS space with a rescaled AdS radius. This result is true regardless of the choice  of gauge. Once we are in AdS space the metric perturbations can be reconstructed using the holographic gauge HKLL smearing functions \cite{Kabat:2012hp} and then one can perform a reverse Weyl rescaling to go back to the original half-flat space.

Our work can be further extended and generalized in several directions thus providing interesting avenues for future investigations. The first and perhaps most natural check our proposal is to verify the bulk reconstruction at the level of two point functions for the Maxwell field and linearized metric perturbations using the CFT$_3$ two point functions. We only checked the bulk reconstruction of Maxwell and linearized metric perturbations here at the one-point function level. Although this might appear tedious (more components, gauge-fixing etc.), conceptually it is no more challenging that the reconstruction of the bulk two-point functions for the conformally coupled (massless) scalar fields. The second straightforward generalization one can do is the scalar bulk reconstruction for $d\geq3$ using the appropriate generalized smearing functions \cite{Aoki:2023lgr, Aoki:2021ekk, DelGrosso:2019gow}. We leave this for a follow up work. The third and perhaps less obvious exercise is to extend our construction to massive bulk scalars. In this case our intuition is that we will have to start with a massive scalar in Poincar\'e AdS with a suitable spacetime-dependent mass-terms, so that after performing a Weyl rescaling we end up with a normal massive scalar in (half)-flat spacetime. We leave this construction for a follow up work. Next, regarding the case of Maxwell and linerized metric fluctuations, we restricted ourselves to $3+1$-dimensions whereby we were aided by certain dimensional advantages. It would be interesting to extend this construction to arbitrary dimensions. For example, we know that in higher or general dimensions the Maxwell equations are not conformal invariant, so starting from a Maxwell equations in (half-)flat space would lead us to Maxwell equation in Poincar\'e AdS with some extra terms (i.e. inhomogeneous Maxwell equations with spacetime dependent sources). However the important point is that the equation would still be \emph{linear} in the fields and one can use a spacelike Green's function to reconstruct this Maxwell field ! Then in principle, one can use the holographic gauge smearing function to reconstruct the complimentary function part of the solution, and the spacelike AdS Green's function term as a particular integral. Once we have this full CFT$_d$ solution to the inhomogeneous Maxwell equation in Poincar\'e AdS$_{d+1}$ spacetime, we can perform the inverse Weyl transformation back to half-flat spacetime, thereby accomplishing the bulk reconstruction at leading order. A similar trick can be applied for reconstructing linearized metric perturbations of the Einstein field equations in arbitrary $d+1$-dimensional half-flat spacetimes in terms of Einstein field equations with a source in Poincar\'e AdS space. We leave the task of reconstructing Maxwell and Fierz-Pauli field in general ($d+1$-) dimensional half-flat space   for a follow up work. Another interesting direction to extend our work is to attempt $1/N$ corrections, i.e. reconstructing interacting fields in half-flat space. In this work we restricted ourselves to the free field equations or equivalently infinite $N$, but one would certainly need to go beyond the free field limit or equivalently CFT 2-point functions as was performed in the AdS case \cite{Kabat:2011rz, Kabat:2012hp, Kabat:2013wga} to gain more insight into gravitational dressings in flat spacetime. To this end, we believe that the method of Kabat, Lifschytz and Lowe of using spacelike Green's functions and higher dimensional multi-trace CFT oeprators can exported with suitable minor alterations (owing to the Weyl rescaling transformations) to the (half-)flat spacetime. These alterations would entail, as alluded to earlier, starting with the AdS couplings which are spacetime dependent to offset the effects of Weyl rescaling transformations from half-flat space to AdS space (see \cite{Caputa:2017yrh} for inclusion of such spacetime dependent quartic scalar couplings and mass under Weyl rescalings albeit in a different context). Another interesting direction where this work can be extended is by including fermions. In this work, we have exclusively worked with bosonic fields, however one also needs to understand reconstruction of fermions coupled to (quantum) gravity in holographic set-ups e.g. \cite{Foit:2019nsr}.  Finally, perhaps the most crucial issue would be to extend our results to the full Minkowski space instead of the half-Minkowski space\footnote{Recently some authors have explored holographic in bulk spacetimes with a physical boundary, e.g. in \cite{Kawamoto:2023nki}, the authors consider holography in half-de Sitter spacetime. However our half-flat space holographic reconstruction has no relation to the half-de Sitter holography}. To this end, recall that in the half-flat spacetime, the bulk fields are required to vanish at the holographic screen $z=0$, as a simple consequence of the normalizable fall behavior of quantum fields at the conformal boundary (again $z=0$), in Poincar\'e AdS. Thus, if one would consider quantum fields in Poinca\'e AdS with more general (dynamical) boundary conditions (see e.g \cite{Krishnan:2016dgy}) at the conformal boundary. With such dynamical boundary conditions on the fields, if one performs a Weyl transformation to half-flat space, one expects to get rid of the rigid or vanishing (Dirichlet) boundary conditions at the $z=0$ hyperplane in the Weyl transformed half-flat space. Then by continuity one can extend the solution beyond the $z=0$, i.e. to the $z< 0$ half of flat space. Although this is a bit of a long shot, if such an exercise is pursued to fruition, it would truly lead to a holographic formulation of quantum gravity in flat space in terms of a more garden variety relativistic CFT with one less space dimension (as opposed to a Carrollian or Celestial CFT) such as $\mathcal{N}=4$ Super-Yang-Mills theory.

\section*{Acknowledgements}
%Authors thank (....) for useful comments on the draft. 
We thank Dan Kabat for his perceptive comments on an early version of the draft. The work of SR is supported by the IIT Hyderabad seed grant SG/IITH/F171/2016-17/SG-47. A.B is supported by Mathematical Research Impact Centric Support Grant (MTR/2021/000490) by the Department of Science and Technology Science and Engineering Research Board (India) and Relevant Research Project grant (202011BRE03RP06633-BRNS) by the Board Of Research In Nuclear Sciences (BRNS), Department of Atomic Energy (DAE), India. AB  \& SR thank the speakers and participants of the workshop ``Quantum Information in QFT and AdS/CFT-III" organized at IIT Hyderabad and funded by SERB through a Seminar Symposia (SSY) grant (SSY/2022/000446) and  ``Quantum Information Theory in Quantum Field Theory and Cosmology" between 4-9th June, 2023 hosted by Banff International Research Centre at Canada for useful discussions for useful discussions.

\appendix
\section{A quick derivation of the HKLL smearing function} \label{HKLL redone}
An AdS-invariant function of two points $\left(x,z\right)$ and $(x',z')$
must be of the form $f(\sigma)$ where $\sigma=\frac{\left(x-x'\right)^{2}+z^{2}+z'^{2}}{2\,z\,z'}$
is the AdS-invariant distance function (chordal distance). The action of the Klein-Gordon operator on such a function, i.e.
\[
\left(\square-m^{2}\right)f(\sigma)=0
\]
 is equivalent to the ordinary second order differential equation\footnote{For this we have used the simplification,
\begin{align*}
\Box f(\sigma) & =\frac{1}{\sqrt{-g}}\partial_{M}\left(\sqrt{-g}\,g^{MN}\partial_{N}f(\sigma)\right)\\
 & =\frac{1}{\sqrt{-g}}\partial_{M}\left(\sqrt{-g}\,g^{MN}\partial_{N}\sigma\:f'(\sigma)\right)\\
 & =\frac{1}{\sqrt{-g}}\left(\sqrt{-g}\,g^{MN}\partial_{N}\sigma\:\partial_{M}f'(\sigma)\right)+\frac{1}{\sqrt{-g}}\partial_{M}\left(\sqrt{-g}\,g^{MN}\:\partial_{N}\sigma\right)f'(\sigma)\\
 & =\left(g^{MN}\:\partial_{N}\sigma\:\partial_{M}\sigma\right)\:f''(\sigma)+\frac{1}{\sqrt{-g}}\partial_{M}\left(\sqrt{-g}\,g^{MN}\:\partial_{N}\sigma\right)f'(\sigma)\\
 & =\left[\frac{\sigma^{2}-1}{R^{2}}\:\frac{d^{2}}{d\sigma^{2}}+\left(\frac{d+1}{R^{2}}\right)\,\sigma\:\frac{d}{d\sigma}\right]f(\sigma).
\end{align*}
},
\begin{equation}
\left[\left(\sigma^{2}-1\right)\frac{d^{2}}{d\sigma^{2}}+\left(d+1\right)\,\sigma\:\frac{d}{d\sigma}-\Delta\left(\Delta-d\right)\right]f(\sigma)=0,\label{eq: box on bilocal}
\end{equation}
where $\Delta \equiv \frac{d}{2}+\sqrt{\left(\frac{d}{2}\right)^2+m^2 R^2}$. This equation can be easily solved
with the appropriate bc's to yield Wightman functions for a scalar
field and so on. Now, if we want to compute the smearing function,
one needs to take the limit, $z'\rightarrow0$ in which case $\sigma$
becomes ill-defined, $\lim_{z'\rightarrow0}\sigma\rightarrow\infty$.
So in this limit one needs to consider the combination, $\sigma z'$,
and express the smearing function in terms of this well defined quantity: $\sigma z'$. In this limit, in terms of $\sigma z'$ variable the equation \eqref{eq: box on bilocal} becomes,
\begin{equation}
\left[\left(\sigma z'\right)^{2}\frac{d^{2}}{d\left(\sigma z'\right)^{2}}+\left(d+1\right)\left(\sigma z'\right) \:\frac{d}{d\left(\sigma z'\right)}-\Delta\left(\Delta-d\right)\right]K(\sigma z')=0. \label{box on smearing}
\end{equation}
This is a homogeneous equation, and the solution will be a monomial.
For spacelike support i.e. for $\sigma>1$ or equivalently, $\sigma z'>0$ in the limit $z'\rightarrow 0$, we need to include a factor of $\theta\left(\sigma z'\right)$. Thus we arrive at the following ansatz for the
smearing function,
\[
K\left(\sigma z'\right)=\left(\sigma z'\right)^{\alpha}\:\theta(\sigma z').
\]
for positive $\alpha$\footnote{If $\alpha$ assumes negative values, the action of derivatives on $K(\sigma z')$ produces singularities and the equation \eqref{box on smearing} is not satisfied.}. Inserting this ansatz and solving for $\alpha$ gives us, 
\[
\alpha=\Delta-d,-\Delta.
\]
Choosing the positive root, we arrive at the HKLL smearing function expression in Poincar\'{e} chart,
\[
K\left(\sigma z'\right)=\left(\sigma z'\right)^{\Delta-d}\:\theta(\sigma z').
\]

\section{Lorentz invariant form of the smearing function}\label{flat space EOM check}
This CFT to half-flat space smearing function for conformally coupled scalar fields, \eqref{half flat smearing},
can be expressed in Lorentz-invariant form as,
\begin{equation}
K\left(s\right)=2^{\frac{d-1}{2}}s^{-\frac{d-1}{2}}\:\theta(s)\label{eq: flat space HKLL for conformally coupled fields}
\end{equation}
where $s=z^{2}-\left(t-t'\right)^{2}-\left(\boldsymbol{y}-\boldsymbol{y}'\right)^{2}$
is the Lorentz invariant interval \emph{Wick-rotated} in the $\boldsymbol{y}$
directions i.e. $\boldsymbol{y}=i\boldsymbol{x}$. Here we directly
check that this smearing function satisfies the equation of motion
for conformally coupled fields, namely
\[
\Box\:K(s)=0\,.
\]
To this end, we use the form of the $\Box$ acting on any function
of the Lorentz-invariant interval,
\begin{align}
\Box\:K(s) & =\eta^{MN}\frac{\partial}{\partial x^{M}}\frac{\partial}{\partial x^{N}}\:K(s)\,,\nonumber \\
 & =\eta^{MN}\frac{\partial}{\partial x^{M}}\left(\frac{\partial s}{\partial x^{N}}\:K'(s)\right)\,,\nonumber \\
 & =\eta^{MN}\frac{\partial^{2}s}{\partial x^{M}\partial x^{N}}\:K'(s)+\eta^{MN}\frac{\partial s}{\partial x^{M}}\frac{\partial s}{\partial x^{N}}\:K"(s)\,.\label{eq: form of Box for Lorentz invariant}
\end{align}
In the Wick-rotated signature, $\eta^{MN}=\left(+,--\ldots\right)$.
So we get,
\[
\eta^{MN}\frac{\partial^{2}s}{\partial x^{M}\partial x^{N}}=2\left(d+1\right),\qquad\eta^{MN}\frac{\partial s}{\partial x^{M}}\frac{\partial s}{\partial x^{N}}=4s.
\]
Plugging this in (\ref{eq: form of Box for Lorentz invariant}), we
obtain the desired result:
\begin{align*}
    \Box\:K(s) & =2\left(d+1\right)\:K'(s)+4s\:K''(s)\,,\\
 & =2\left(d+1\right)\:\frac{d}{ds}\left(2^{\frac{d-1}{2}}s^{-\frac{d-1}{2}}\:\theta(s)\right)+4s\:\frac{d^{2}}{ds^{2}}\left(2^{\frac{d-1}{2}}s^{-\frac{d-1}{2}}\:\theta(s)\right)\,,\\
 & =2^{\frac{d+1}{2}}\left(d+1\right)\left[-\frac{d-1}{2}s^{-\frac{d+1}{2}}\:\theta(s)+s^{-\frac{d-1}{2}}\:\delta(s)\right]+2^{\frac{d+3}{2}}s\:\frac{d}{ds}\left[-\frac{d-1}{2}s^{-\frac{d+1}{2}}\:\theta(s)+s^{-\frac{d-1}{2}}\:\delta(s)\right]\,,\\
 & =2^{\frac{d+1}{2}}\left(d+1\right)\left[-\frac{d-1}{2}s^{-\frac{d+1}{2}}\:\theta(s)+s^{-\frac{d-1}{2}}\:\delta(s)\right]\\
 & \qquad\qquad\qquad\qquad+2^{\frac{d+3}{2}}s\:\left[\left(\frac{d-1}{2}\right)\left(\frac{d+1}{2}\right)s^{-\frac{d+3}{2}}\:\theta(s)-\left(d-1\right)s^{-\frac{d+1}{2}}\:\delta(s)+s^{-\frac{d-1}{2}}\:\delta'(s)\right]\,,\\
 & =\left[-2^{\frac{d-1}{2}}\left(d+1\right)\left(d-1\right)s^{-\frac{d+1}{2}}\:\theta(s)+2^{\frac{d+1}{2}}\left(d+1\right)s^{-\frac{d-1}{2}}\:\delta(s)\right]\\
 & \qquad\qquad\qquad\qquad+\:\left[2^{\frac{d-1}{2}}\left(d+1\right)\left(d-1\right)s^{-\frac{d+1}{2}}\:\theta(s)-2^{\frac{d+3}{2}}\left(d-1\right)s^{-\frac{d+1}{2}}\:\delta(s)+2^{\frac{d+3}{2}}s^{-\frac{d-3}{2}}\:\delta'(s)\right]\,,\\
 & =2^{\frac{d+1}{2}}\left(3-d\right)s^{-\frac{d-1}{2}}\:\delta(s)+2^{\frac{d+3}{2}}s^{-\frac{d-3}{2}}\:\delta'(s)\,,\\
 & =2^{\frac{d+1}{2}}\left(3-d\right)s^{-\frac{d-1}{2}}\:\delta(s)-2^{\frac{d+3}{2}}\frac{d}{ds}\left(s^{-\frac{d-3}{2}}\right)\:\delta(s)\,,\\
 & =2^{\frac{d+1}{2}}\left(3-d\right)s^{-\frac{d-1}{2}}\:\delta(s)+2^{\frac{d+1}{2}}\left(d-3\right)s^{-\frac{d-1}{2}}\:\delta(s)\,,\\
 & =0.
 \end{align*}
\section{Bulk SUGRA Wightman function} \label{bulk Wightman}
For half-Minkowski space, the normalized mode functions are,
\begin{equation}
f_{\boldsymbol{k}}(X)=\frac{1}{\sqrt{\left(2\pi\right)^{d-1}\pi\omega_{\boldsymbol{k}}}}e^{-i\omega_{\boldsymbol{k}}t+i\boldsymbol{k}\cdot\boldsymbol{x}}\sin k_{z}z,\qquad\boldsymbol{k}=\left(k_{1},k_{2},\ldots,k_{d-1}\right)
\end{equation}
Specializing to the case where $X,Y$ are spacelike separated, we
can choose for convenience, 
\[
X=(0,\boldsymbol{0},z_{1}),\quad Y=(0,\boldsymbol{0},z_{2}).
\]
Then the Wightman function expression reads,
\begin{align*}
\Delta_{+}(X,Y) & =\frac{1}{\left(2\pi\right)^{d}}\int_{-\infty}^{\infty}d^{d-1}\boldsymbol{k}\:\int_{-\infty}^{\infty}dk_{z}\,\frac{\sin k_{z}z_{1}\:\sin k_{z}z_{2}}{\sqrt{\boldsymbol{k}^{2}+k_{z}^{2}}}.\\
 & =\frac{1}{\left(2\pi\right)^{d}}\int\frac{d^{d}\boldsymbol{k}}{\left|\boldsymbol{k}\right|}\,\sin k_{z}z_{1}\:\sin k_{z}z_{2}.
\end{align*}
Note that now $\boldsymbol{k}$ is a $d$-dimensional euclidean vector.
Next we massage this expression so that it looks exactly like the
difference of a pair of Wightman functions in full Minkowski space,
\begin{align*}
\Delta_{+}(X,Y) & =\frac{1}{2\left(2\pi\right)^{d}}\int\frac{d^{d}\boldsymbol{k}}{\left|\boldsymbol{k}\right|}\,\cos k_{z}\left(z_{1}-z_{2}\right)-\cos k_{z}\left(z_{1}+z_{2}\right)\\
 & =\frac{1}{4\left(2\pi\right)^{d}}\int\frac{d^{d}\boldsymbol{k}}{\left|\boldsymbol{k}\right|}\,\left(e^{ik_{z}\left(z_{1}-z_{2}\right)}+e^{-ik_{z}\left(z_{1}-z_{2}\right)}-e^{ik_{z}\left(z_{1}+z_{2}\right)}-e^{-ik_{z}\left(z_{1}+z_{2}\right)}\right)\\
 & =\frac{1}{4\left(2\pi\right)^{d}}\int\frac{d^{d}\boldsymbol{k}}{\left|\boldsymbol{k}\right|}\,\left(e^{ik_{z}\left(z_{1}-z_{2}\right)}-e^{ik_{z}\left(z_{1}+z_{2}\right)}\right)+\frac{1}{4\left(2\pi\right)^{d}}\underbrace{\int\frac{d^{d}\boldsymbol{k}}{\left|\boldsymbol{k}\right|}\,\left(e^{-ik_{z}\left(z_{1}-z_{2}\right)}-e^{-ik_{z}\left(z_{1}+z_{2}\right)}\right)}_{\boldsymbol{k}\rightarrow-\boldsymbol{k}}\\
 & =\frac{1}{2\left(2\pi\right)^{d}}\int\frac{d^{d}\boldsymbol{k}}{\left|\boldsymbol{k}\right|}\,e^{ik_{z}\left(z_{1}-z_{2}\right)}-\frac{1}{2\left(2\pi\right)^{d}}\int\frac{d^{d}\boldsymbol{k}}{\left|\boldsymbol{k}\right|}\,e^{ik_{z}\left(z_{1}+z_{2}\right)}\\
 & =D_{+}(X,Y)-D_{+}(X,Y_{*})
\end{align*}
where $Y_{*}=\left(0,0,-z_{2}\right)$ is the location of insertion
of an ``image'' operator and $D_{+}$ is the Wightman function for
massless scalar fields in full Minkowski space. Next we compute $D_{+}(X,Y)$.
\[
D_{+}(X,Y)=\frac{1}{2\left(2\pi\right)^{d}}\int\frac{d^{d}\boldsymbol{k}}{\left|\boldsymbol{k}\right|}\,e^{ik_{z}\left(z_{1}-z_{2}\right)}\,.
\]
Going over to polar coordinates, $\left(q,\theta,\boldsymbol{\chi}\right),\boldsymbol{\chi}\in S_{d-2}$,
in $\boldsymbol{k}$-space,
\[
D_{+}(X,Y)=\frac{\Omega_{d-2}}{2\left(2\pi\right)^{d}}\int_{0}^{\infty}dq\:q^{d-2}\:\int_{0}^{\pi}d\theta\left(\sin\theta\right)^{d-2}\:e^{iq\left|z_{1}-z_{2}\right|\cos\theta}
\]
where $\Omega_{n}=\frac{2\pi^{n/2}}{\Gamma(n/2)}$ is the volume of the $n$-sphere $(S_{n})$.
 Using the identity,
\[
\int_{0}^{\pi}e^{i\alpha\cos\theta}\sin^{2m}\theta\,d\theta=\sqrt{\pi}\:2^{m}\:\Gamma\left(m+\frac{1}{2}\right)\:\left|\alpha\right|^{-m}\:|J_{m}\left(\left|\alpha\right|\right)
\]
we get,
\[
D_{+}(X,Y)=\frac{\Omega_{d-2}\:\sqrt{\pi}\:2^{m}\:\Gamma\left(\frac{d-1}{2}\right)}{2\left(2\pi\right)^{d}\left|z_{1}-z_{2}\right|^{\frac{d-2}{2}}}\int_{0}^{\infty}dq\:q^{\frac{d-2}{2}}\:J_{\frac{d-2}{2}}\left(q\left|z_{1}-z_{2}\right|\right).
\]
Next, defining, $y=q\left|z_{1}-z_{2}\right|$, we obtain
\begin{align}
D_{+}(X,Y) & =\frac{\Omega_{d-2}2^{\frac{d-2}{2}}\sqrt{\pi}\Gamma\left(\frac{d-1}{2}\right)}{2\left(2\pi\right)^{d}\left|z_{1}-z_{2}\right|^{d-1}}\int_{0}^{\infty}dy\:y^{\frac{d-2}{2}}\:J_{\frac{d-2}{2}}\left(y\right)\,,\nonumber \\
 & =\frac{\Omega_{d-2}\:2^{d-3}\:\Gamma^{2}\left(\frac{d-1}{2}\right)}{\left(2\pi\right)^{d}}\frac{1}{\left|z_{1}-z_{2}\right|^{d-1}}.\label{eq: Wightman function in arbitrary dimensions}
\end{align}
 For $d=2$,
\[
D_{+}(X,Y)=\frac{1}{4\pi}\frac{1}{\left|z_{1}-z_{2}\right|}
\]
and for $d=3$, we get,
\[
D_{+}(X,Y)=\frac{1}{4\pi^{2}}\frac{1}{\left|z_{1}-z_{2}\right|^{2}},
\]
which can be also be obtained by taking a massless limit ($m\rightarrow0$)
of the Wightman function of a massive scalar for spacelike separations,
namely $\Delta_{+}(r)=\frac{m}{4\pi^{2}r}K_{1}\left(m\,r\right)$.
So finally for spacelike separated points in the half-Minkowski space,
we get the Wightman function,
\begin{equation}
\Delta_{+}\left(\left.0,\boldsymbol{0},z_{1}\right|0,\boldsymbol{0},z_{2}\right)=\frac{\Omega_{d-2}\:2^{d-3}\:\Gamma^{2}\left(\frac{d-1}{2}\right)}{\left(2\pi\right)^{d}}\left[\frac{1}{\left|z_{1}-z_{2}\right|^{d-1}}-\frac{1}{\left|z_{1}+z_{2}\right|^{d-1}}\right].\label{eq: Wightman function for z-separated points in 1/2 Minkowski}
\end{equation}
\\

For timelike separations we can choose, $X=\left(t,\boldsymbol{0},z\right),Y=\left(0,\boldsymbol{0},z\right)$.
Then the Wightman function expression reads,
\begin{align*}
\Delta_{+}(X,Y) & =\frac{1}{\left(2\pi\right)^{d}}\int_{-\infty}^{\infty}\frac{d^{d}\boldsymbol{k}}{\left|\boldsymbol{k}\right|}\,e^{-i\left|\boldsymbol{k}\right|t}\:\sin^{2}k_{z}z\,,\\
 & =\frac{1}{2\left(2\pi\right)^{d}}\int_{-\infty}^{\infty}\frac{d^{d}\boldsymbol{k}}{\left|\boldsymbol{k}\right|}\,e^{-i\left|\boldsymbol{k}\right|t}\:\left(1-\cos2k_{z}z\right)\,,\\
 & =\frac{1}{2\left(2\pi\right)^{d}}\int_{-\infty}^{\infty}\frac{d^{d}\boldsymbol{k}}{\left|\boldsymbol{k}\right|}\,e^{-i\left|\boldsymbol{k}\right|t}\:\left(\frac{2-e^{i\:2k_{z}z}-e^{-i\:2k_{z}z}}{2}\right)\,,\\
 & =\frac{1}{2\left(2\pi\right)^{d}}\int_{-\infty}^{\infty}\frac{d^{d}\boldsymbol{k}}{\left|\boldsymbol{k}\right|}\,e^{-i\left|\boldsymbol{k}\right|t}\:\left(1-e^{i\:2k_{z}z}\right)\,,\\
 & =\frac{1}{2\left(2\pi\right)^{d}}\int_{-\infty}^{\infty}\frac{d^{d}\boldsymbol{k}}{\left|\boldsymbol{k}\right|}\,e^{-i\left|\boldsymbol{k}\right|t}-\frac{1}{2\left(2\pi\right)^{d}}\int_{-\infty}^{\infty}\frac{d^{d}\boldsymbol{k}}{\left|\boldsymbol{k}\right|}\,e^{-i\left(\left|\boldsymbol{k}\right|t-k_{z}(2\,z)\right)}\,,\\
 & =D_{+}(X,Y)-D_{+}(X,Y_{*})\,,
\end{align*}
where $Y_{*}=\left(0,0,-z\right)$ is the location of insertion of
an ``image'' operator and $D_{+}$ is the Wightman function for
massless scalar fields in full Minkowski space. Next we compute $D_{+}(X,Y)$.
\[
D_{+}(X,Y)=\frac{1}{2\left(2\pi\right)^{d}}\int_{-\infty}^{\infty}\frac{d^{d}\boldsymbol{k}}{\left|\boldsymbol{k}\right|}\,e^{-i\left|\boldsymbol{k}\right|t}\,.
\]
Switching to polar coordinates,
\begin{align*}
D_{+}(t) & =\frac{\Omega_{d-1}}{2\left(2\pi\right)^{d}}\int_{0}^{\infty}dq\,q^{d-2}\,e^{-iq\left(t-i\varepsilon\right)}\,,\\
 & =\frac{\Omega_{d-1}}{2\left(2\pi\right)^{d}}\left(i\frac{d}{dt}\right)^{d-2}\left[\int_{0}^{\infty}dq\,e^{-iq\left(t-i\varepsilon\right)}\right]\,,\\
 & =\frac{\Omega_{d-1}}{2\left(2\pi\right)^{d}}\left(i\frac{d}{dt}\right)^{d-2}\left[\frac{1}{i\left(t-i\varepsilon\right)}\right]\,,\\
 & =\frac{\Omega_{d-1}\:\Gamma\left(d-1\right)\,(-)^{d-2}i^{d-2}}{2\left(2\pi\right)^{d}i}\frac{1}{\left(t-i\varepsilon\right)^{d-1}}\,,\\
 & =\frac{\Omega_{d-1}\:\Gamma\left(d-1\right)}{2\left(2\pi\right)^{d}i^{d-1}}\frac{1}{\left(t-i\varepsilon\right)^{d-1}}\,,\\
 & =\frac{\Omega_{d-2}\:2^{d-3}\,\Gamma^{2}\left(\frac{d-1}{2}\right)}{\left(2\pi\right)^{d}i^{d-1}}\frac{1}{\left(t-i\varepsilon\right)^{d-1}}.
\end{align*}
Here in the last step, we have used the identity,
\begin{equation}
\frac{\sqrt{\pi}}{2}\frac{\Gamma\left(\frac{d-1}{2}\right)}{\Gamma\left(\frac{d}{2}\right)}\Gamma(d-1)=2^{d-3}\,\Gamma^{2}\left(\frac{d-1}{2}\right).
\end{equation}

Next we compute,
\[
D_{+}(X,Y_{*})=\frac{1}{2\left(2\pi\right)^{d}}\int_{-\infty}^{\infty}\frac{d^{d}\boldsymbol{k}}{\left|\boldsymbol{k}\right|}\,e^{-i\left(\left|\boldsymbol{k}\right|t-k_{z}(2\,z)\right)}
\]
Switching to polar coordinates we get,
\begin{align*}
D_{+}(X,Y_{*}) & =\frac{\Omega_{d-2}}{2\left(2\pi\right)^{d}}\int_{0}^{\infty}dq\,q^{d-2}e^{-i\,q\,\left(t-i\varepsilon\right)}\int_{0}^{\pi}d\theta\,\sin^{d-2}\theta\:e^{i\,2zq\cos\theta}\,,\\
 & =\frac{\Omega_{d-2}\sqrt{\pi}\:\:\Gamma\left(\frac{d-1}{2}\right)}{2\left(2\pi\right)^{d}\left|z\right|^{\frac{d-2}{2}}}\int_{0}^{\infty}dq\,q^{\frac{d-2}{2}}e^{-i\,q\,\left(t-i\varepsilon\right)}\:J_{\frac{d-2}{2}}\left(2\left|z\right|q\right)\,,\\
 & =\frac{\Omega_{d-2}\sqrt{\pi}\:\:\Gamma\left(\frac{d-1}{2}\right)}{2\left(2\pi\right)^{d}\left|z\right|^{d-1}2^{\frac{d}{2}}}\int_{0}^{\infty}dy\,y^{\frac{d-2}{2}}e^{-i\,y\,\left(\frac{t-i\varepsilon}{2\left|z\right|}\right)}\:J_{\frac{d-2}{2}}\left(y\right)\,,\\
 & =\frac{\Omega_{d-2}\:2^{d-3}\,\Gamma^{2}\left(\frac{d-1}{2}\right)}{\left(2\pi\right)^{d}i^{d-1}}\frac{1}{\left(\left(t-i\varepsilon\right)^{2}-4z^{2}\right)^{\frac{d-1}{2}}}.
\end{align*}
Thus for timelike separations, the Wightman function in half-Minkowski
space is,
\begin{equation}
\Delta_{+}\left(\left.t,\boldsymbol{0},z\right|0,\boldsymbol{0},z\right)=\frac{\Omega_{d-2}\:2^{d-3}\,\Gamma^{2}\left(\frac{d-1}{2}\right)}{\left(2\pi\right)^{d}i^{d-1}}\left[\frac{1}{\left(t-i\varepsilon\right)^{d-1}}-\frac{1}{\left(\left(t-i\varepsilon\right)^{2}-4z^{2}\right)^{\frac{d-1}{2}}}\right].\label{eq: Wightman function for t-separated points in 1/2 Minkowski}
\end{equation}
The $d=1$ case has to be done separately since the expression for general dimensions \eqref{eq: Wightman function in arbitrary dimensions} becomes trivial when $d=1$. For $d=1$, the normalized (positive energy) modes have the form \begin{equation}
    f_k(x,t)=\frac{e^{-i\omega t} \sin(kx)}{\sqrt{\pi\omega}},\,\, k=\omega>0.
\end{equation}
The bulk Wightman function is then,
\begin{align}
\langle \phi (z_1,t_1)\,\,\phi (z_2,t_2)\rangle &= \int_0^\infty\frac{dk}{\sqrt{\pi~k}}\int_0^\infty\frac{dp}{\sqrt{\pi~p}} e^{-i(k~t_1-p~t_2)} \sin{k z_1}\,\sin{p z_2}\, \delta(k-p)\,, \nonumber \\
&=\int_0^\infty\frac{dk}{\pi k} \, e^{-ik(t_1-t_2)}\,\sin{k z_1}\,\sin{k z_2}\,,\nonumber \\
&=\frac{1}{4\pi}\ln\left[\frac{(t_1-t_2)^2-(z_1+z_2)^2}{(t_1-t_2)^2-(z_1-z_2)^2}\right]\,.\label{Wightman} 
\end{align}
Here as usual we need the prescription $t_1-t_2-i\varepsilon$ to define the Wightman function smoothly in the limit when $t_1=t_2$. The half-Minkowski space does not have translation invariance in the $z$-direction, and this is reflected in the dependence on $(z_1+z_2)$. As in the case for $d>0$, the Dirichlet boundary at $z=0$ can be replaced by the full Minkowski space with an image operator placed at $\left(t_2,-z_2\right)$ so that the final answer \eqref{Wightman} is an image sum i.e. sum of the full Minkowski space Wightman functions
\[\langle\phi (z_1,t_1)~\phi(z_2,t_2)\rangle + \langle\phi(z_1,t_1)~\phi(-z_2,t_2)\rangle.
\]
\section{Matching the SUGRA and CFT normalization of the two-point functions} \label{fix C}
The CFT representation of a local bulk operator in half-Minkowski
space is
\[
\varphi(\boldsymbol{x},z)=C_{d}\left(\frac{l}{z}\right)^{\frac{d-1}{2}}\int_{\left|\boldsymbol{y}-\boldsymbol{x}\right|<z}d^{d}\boldsymbol{y}\:\left(\frac{z^{2}-\left(\boldsymbol{y-x}\right)^{2}}{2z}\right)^{\Delta-d}\:\mathcal{O}\left(\boldsymbol{y}\right)
\]
Switching to spherical polar coordinates,
\[
\boldsymbol{y-x}=\rho\:\boldsymbol{\chi},\quad\boldsymbol{\chi}\cdot\boldsymbol{\chi}=1,
\]
where the points $\boldsymbol{\chi}$ lies on $S_{d-1}$, we get,
\[
\varphi(\boldsymbol{x},z)=C_{d}\left(\frac{l}{z}\right)^{\frac{d-1}{2}}\int_{0}^{z}d\rho\:\rho^{d-1}\:\left(\frac{z^{2}-\rho^{2}}{2z}\right)^{\Delta-d}\int d\Omega_{d-1}\:\mathcal{O}\left(\boldsymbol{x}+\rho\boldsymbol{\chi}\right)\,.
\]
Next we define, $\rho=zw$ and get,
\begin{align*}
\varphi(\boldsymbol{x},z) & =C_{d}\left(\frac{l}{z}\right)^{\frac{d-1}{2}}\frac{z^{d+2(\Delta-d)}}{\left(2z\right)^{\Delta-d}}\int_{0}^{1}dw\:w^{d-1}\left(1-w^{2}\right)^{\Delta-d}\int d\Omega_{d-1}\:\mathcal{O}\left(\boldsymbol{x}+z\:w\boldsymbol{\chi}\right)\,,\\
 & =\frac{C_{d}\left(l\right)^{\frac{d-1}{2}}}{2^{\Delta-d}}z^{\Delta-\frac{d-1}{2}}\int_{0}^{1}dw\:w^{d-1}\left(1-w^{2}\right)^{\Delta-d}\int d\Omega_{d-1}\:\mathcal{O}\left(\boldsymbol{x}+z\:w\boldsymbol{\chi}\right)\,.
\end{align*}
\\
The only place where $z$ appears now is in the integrand of the angular
integral. We expand the angular integrand around $z\rightarrow0$
and keep only the leading term,
\[
\mathcal{O}\left(x+z\:w\boldsymbol{\chi}\right)\approx\mathcal{O}(\boldsymbol{x}).
\]
In this limit (extrapolate limit), the SUGRA field works out to be,
\begin{align*}
\varphi(\boldsymbol{x},z) & \sim\frac{C_{d}\left(l\right)^{\frac{d-1}{2}}}{2^{\Delta-d}}\underbrace{\left(\int_{0}^{1}dw\:w^{d-1}\left(1-w^{2}\right)^{\Delta-d}\right)}_{\frac{1}{2}B\left(\frac{d}{2},\Delta-d+1\right)}\underbrace{\left(\int d\Omega_{d-1}\right)}_{\Omega_{d-1}}\:z^{\Delta-\frac{d-1}{2}}\:\mathcal{O}\left(\boldsymbol{x}\right)\,,\\
 & \sim\frac{C_{d}\left(l\right)^{\frac{d-1}{2}}\Omega_{d-1}\Gamma\left(\frac{d}{2}\right)\Gamma\left(\Delta-d+1\right)}{2^{\Delta-d+1}\Gamma\left(\Delta-\frac{d}{2}+1\right)}\:z^{\Delta-\frac{d-1}{2}}\:\mathcal{O}\left(\boldsymbol{x}\right)\,.
\end{align*}
Finally we put in, 
\[
\Delta=\frac{d+1}{2}
\]
and obtain,
\begin{equation}
\varphi(\boldsymbol{x},z)\sim\frac{C_{d}l^{\frac{d-1}{2}}\Omega_{d-1}\Gamma\left(\frac{d}{2}\right)\Gamma\left(\frac{3-d}{2}\right)}{2^{\frac{3-d}{2}}\:\Gamma\left(\frac{3}{2}\right)}\:z\:\mathcal{O}\left(\boldsymbol{x}\right).\label{eq: z=00003D0 limit}
\end{equation}
\[
\]
For $d=1$, one has $\Delta=1$ and we get
\begin{align*}
\varphi(t,z) & \sim\frac{C_{1}\Omega_{0}\Gamma\left(\frac{1}{2}\right)\Gamma\left(1\right)}{2\:\Gamma\left(\frac{3}{2}\right)}\:z\:\mathcal{O}\left(t\right)\\
 & \sim2C_{1}\:z\:\mathcal{O}\left(t\right).
\end{align*}
So the $z\rightarrow0$ limit of two point function is,
\begin{align}
\langle\varphi(t_{1},z_{1})\varphi(t_{2},z_{2})\rangle & \sim4C_{1}^{2}\;z^{2}\:\langle\mathcal{O}\left(t_{1}\right)\mathcal{O}\left(t_{2}\right)\rangle\,,\nonumber \\
 & \sim4C_{1}^{2}\;z^{2}\:\frac{1}{\left(t_{1}-t_{2}\right)^{2}}\label{eq: extrapolate 2 point smearing}\,.
\end{align}
In particular for $z_{1}=z_{2}=z$ and $t_{1}=t,t_{2}=0$, 
\[
\langle\varphi(t,z)\;\varphi(0,z)\rangle\sim4C_{1}^{2}\;z^{2}\:\frac{1}{t^{2}}\,.
\]
From the direct SUGRA calculation, we found,
\begin{align}
\langle\varphi(t,z)\;\varphi(0,z)\rangle & \sim\lim_{z\rightarrow0}\frac{1}{4\pi}\ln\left(\frac{t^{2}}{t^{2}-4z^{2}}\right)\,,\nonumber \\
 & \sim\frac{1}{\pi}\frac{z^{2}}{t^{2}}\,.\label{eq: extrapolate 2 point bulk}
\end{align}
Comparing the two forms (\ref{eq: extrapolate 2 point smearing},
\ref{eq: extrapolate 2 point bulk}) we get,
\[
4C_{1}^{2}=\frac{1}{\pi}\Rightarrow C_{1}=\frac{1}{2\sqrt{\pi}}.
\]
Next, for $d=2$ (i.e. $\Delta=\frac{3}{2}$)
\begin{align*}
\varphi(\boldsymbol{x},z) & \sim\frac{C_{2}\:l^{\frac{1}{2}}\Omega_{1}\Gamma\left(1\right)\Gamma\left(\frac{1}{2}\right)}{2^{\frac{1}{2}}\:\Gamma\left(\frac{3}{2}\right)}\:z\:\mathcal{O}\left(\boldsymbol{x}\right)\,,\\
 & \sim C_{2}\:l^{\frac{1}{2}}2^{\frac{3}{2}}\pi\:z\:\mathcal{O}\left(\boldsymbol{x}\right)\,.
\end{align*}
So, the $z=0$ limit of the SUGRA two-point function (setting $l=1$)
is,
\[
\langle\varphi(\boldsymbol{x}_{1},z_{1})\varphi(\boldsymbol{x}_{2},z_{2})\rangle\sim C_{2}^{2}\:8\pi^{2}\:z_1 z_2\:\langle\mathcal{O}\left(\boldsymbol{x}_{1}\right)\:\mathcal{O}\left(\boldsymbol{x}_{2}\right)\rangle\,.
\]
In particular for $\boldsymbol{x}_{1}=\left(t,0\right)$ and $\boldsymbol{x}_{2}=\left(0,0\right)$
and $z_{1}=z_{2}=z$ we get,
\begin{align}
\langle\varphi(t,0,z)\varphi(0,0,z)\rangle & \sim C_{2}^{2}\;8\pi^{2}\:z^{2}\:\langle\mathcal{O}\left(t,0\right)\:\mathcal{O}\left(0,0\right)\rangle\,,\nonumber \\
 & \sim C_{2}^{2}\:8\pi^{2}\:\frac{z^{2}}{\left(-t^{2}\right)^{3/2}}.\label{eq: sugra 2 point limit from smearing}
\end{align}
From the direct bulk calculation we find,
\begin{align}
\langle\varphi(t,0,z)\varphi(0,0,z)\rangle & \sim\lim_{z\rightarrow0}\frac{i}{4\pi}\left(\frac{1}{\sqrt{t^{2}-4z^{2}}}-\frac{1}{t}\right)\,,\nonumber \\
 & \sim\frac{i}{4\pi}\left(\frac{1}{t}\left(1+\frac{2z^{2}}{t^{2}}\right)-\frac{1}{t}\right)\,,\nonumber \\
 & \sim\frac{1}{2\pi}\:\frac{z^{2}}{\left(-t^{2}\right)^{3/2}}.\label{eq: sugra 2 point limit direct bulk}
\end{align}
Comparing the two asymptotic forms (\ref{eq: sugra 2 point limit from smearing},
\ref{eq: sugra 2 point limit direct bulk}), we get
\[
C_{2}^{2}\:8\pi^{2}=\frac{1}{2\pi}
\]
or,
\[
C_{2}=\frac{1}{4\,\pi^{3/2}}.
\]

More generally, for arbitrary $d$, direct bulk calculation gives
\[
\left\langle \varphi(t,\boldsymbol{0},z)\varphi(0,\boldsymbol{0},z)\right\rangle =\frac{\Omega_{d-2}\:2^{d-3}\,\Gamma^{2}\left(\frac{d-1}{2}\right)}{\left(2\pi\right)^{d}i^{d-1}}\left[\frac{1}{\left(t-i\varepsilon\right)^{d-1}}-\frac{1}{\left(\left(t-i\varepsilon\right)^{2}-4z^{2}\right)^{\frac{d-1}{2}}}\right],\,d\neq1\,.
\]
Then taking the $z\rightarrow0$ limit, we get,
\[
\lim_{z\rightarrow0}\left\langle \varphi(t,\boldsymbol{0},z)\varphi(0,\boldsymbol{0},z)\right\rangle =\frac{\Omega_{d-2}\:2^{d-2}\,\Gamma^{2}\left(\frac{d-1}{2}\right)(d-1)}{\left(2\pi\right)^{d}}\frac{z^{2}}{\left[-\left(t-i\varepsilon\right)^{2}\right]^{\frac{d+1}{2}}},\,d\neq1\,.
\]
But from the boundary smearing function construction (\ref{eq: z=00003D0 limit}),
\[
\lim_{z\rightarrow0}\left\langle \varphi(t,\boldsymbol{0},z)\varphi(0,\boldsymbol{0},z)\right\rangle =\left(\frac{C_{d}l^{\frac{d-1}{2}}\Omega_{d-1}\Gamma\left(\frac{d}{2}\right)\Gamma\left(\frac{3-d}{2}\right)}{2^{\frac{3-d}{2}}\:\Gamma\left(\frac{3}{2}\right)}\right)^{2}\:\frac{z^{2}}{\left[-\left(t-i\varepsilon\right)^{2}\right]^{\frac{d+1}{2}}},\,d<3\,.
\]
Comparing these two we obtain 
\begin{equation}
C_{d}=\left(\frac{\Omega_{d-2}\:2^{d-2}\,\Gamma^{2}\left(\frac{d-1}{2}\right)(d-1)}{\left(2\pi\right)^{d}}\right)^{1/2}\frac{2^{\frac{3-d}{2}}\:\Gamma\left(\frac{3}{2}\right)}{l^{\frac{d-1}{2}}\Omega_{d-1}\Gamma\left(\frac{d}{2}\right)\Gamma\left(\frac{3-d}{2}\right)},d\neq1.\label{eq: gen d}
\end{equation}
In particular, for $d=2$, one obtains
\[
C_{2}=\left(\frac{1}{2\pi}\right)^{1/2}\frac{1}{l^{\frac{1}{2}}2^{\frac{3}{2}}\pi}=\frac{1}{4\,\pi^{3/2}\,l^{1/2}}.
\]
\bibliographystyle{utphys}
\bibliography{ref}

\end{document}